\documentstyle[12pt]{article}
\def\lsim{\mathrel{\rlap{\raise 2.5pt \hbox{$<$}}\lower 2.5pt}}
\def\gsim{\mathrel{\rlap{\raise 2.5pt \hbox{$>$}}\lower 2.5pt}}
\begin{document}
\bibliographystyle{plain}
\thispagestyle{empty}
 \begin{small}
 \begin{flushright}
 IISc-CTS-6/98\\
 hep-ph/9807356\\
 \end{flushright}
 \end{small}
\begin{center}
{\Large
{\bf  
Effective Lagrangians
in $ 2 \, (+\epsilon) $ Dimensions}}

\vskip 2.0cm

B. Ananthanarayan\\

\bigskip

Centre for Theoretical Studies, \\
Indian Institute of Science, \\
Bangalore 560 012, India.\\

\bigskip

\vskip 2cm

\end{center}

\begin{abstract}
The failure of the the loop expansion and effective lagrangians
in two dimensions, which traditionally hinges on a power counting argument
is considered.  We establish that the book keeping device 
for the loop expansion, a role played by (the reciprocal of) the 
pion-decay constant itself vanishes for $d=2$,  thereby going beyond
the power counting argument.     
We point the connection of our results to the distinct phases of
the candidate for the effective lagrangians, the non-linear sigma model, in 
$d=2+\epsilon$, and eventually for
$d=2$.  In light of our results, we recall some of
the relavant features of the multi-flavor Schwinger and 
large $N_f$ $QCD_2$ as candidates for the underlying theory in $d=2$.
\end{abstract}
\newpage
\setcounter{equation}{0}
\section{Introduction}
It has been argued~\cite{sw1,sw2} that when a physical amplitude from
Feynman diagrams using the most general Lagrangian that
involves the relevant degrees of freedom and satisfies
the assumed symmetries of the theory is calculated, then
this amounts to writing down the most general amplitude
that is consistent with the general principles of relativity,
quantum mechanics and the assumed symmetries.  This
argument (which awaits a formal proof~\cite{sw4}), suitably 
generalized leads to the modern effective field theory
of the low-energy strong interaction physics~\cite{hl1,gl1}  
involving the pseudo-scalar degrees of freedom, allowing
for a loop-expansion. A power counting argument due to Weinberg  
that justifies the use of the effective lagrangians/loop expansion is 
a cornerstone of these techniques~\cite{sw1,sw2}. 
The effective lagrangian at leading order is the non-linear
sigma model with the pion decay constant $F_\pi$ playing
the role of the (reciprocal of) the coupling constant.  Furthermore,
inverse powers of the pion decay constant play the role of
a book keeping device for the loop-expansion.
It has been noted that the power counting argument is general
enough to admit an analysis in dimensions $d\neq 4$, and it follows 
from there that the loop-expansion fails in $d=2$~\cite{hl1}.

A well-known result in $d=2$,
due to Coleman~\cite{sc1} and Mermin and Wagner~\cite{mw1},
is that spontaneous symmetry breaking cannot occur. 
It has been pointed out that one of the original proofs 
provided for the Goldstone mechanism~\cite{gsw1}, breaks down 
for $d=2$~\cite{sc1}, which follows from the
impossibility of defining a massless scalar field
theory~\cite{bs1,asw1,aar} for $d=2$.  The latter result is rigorously true, 
as it is based on the underlying Wightman functions of field theory.   
Also it might be legitimate to take the following point of view:
the Coleman-Mermin-Wager theorem says that there are no Goldstone
bosons in $d=2$ and as a result it is not necessary to speak of
an effective field theory.  In constrast, we have found it more fruitful to
examine the foundations of the programme of effective lagrangians.
Here, we shall show that the impossibility of defining
a massless scalar field in $d=2$ implies that $F_\pi$, the
book keeping device of the loop expansion
vanishes identically, thereby providing a
new insight for the failure of the effective lagrangian
technique in $d=2$, and which is independent of the loop expansion.
This is the main conclusion of the present work.
We note that our proof will be valid only in the case of
non-anomalous currents that interpolate between the vacuum and
the candidates for the one-pion occupation number states.

A careful discussion of all the underlying assumptions of
the program, and an investigation of the manner in which
they breakdown in $d=2$ is presented, and the properties 
of the underlying theory and that of the effective theory 
separately.   In what follows the results do not 
particularly depend on whether the
symmetries are actually the axial vector symmetries, or whether
the Goldstone bosons are actually the pions.  We shall, however,
stick to the standard terminology, since no confusion is likely
to arise.

In Section 2 we discuss the foundations of effective lagrangians
and the power counting argument.   We recount the main assumptions
and examine all the assumptions critically and comment on possible
generalizations.  In Section 3 we revisit the
Goldstone mechanism with emphasis on the causes of its breakdown
for $d=2$, by studying more than one proof of the theorem.  
A central result we present here is that the phenomenological
pion-decay constant has to vanish identically in this case.  
We also discuss the relationship of our result
with the established results for the non-linear sigma model for
$d=2+\epsilon$ and for $d=2$.  In Section 4 we examine the implications of
our results, to some candidates for the underlying theory
such as the Schwinger and large $N_f$ $QCD_2$ models.  In Section 5
we summarize our conclusions.

\setcounter{equation}{0}
\section{Effective Lagrangians and Power Counting}
We consider the case of massless QCD with two quark flavors
with the spontaneous breakdown of chiral $SU(2)_L\times SU(2)_R$
to $SU(2)_V$.  The Goldstone theorem then requires that
the following identity be verified:
\begin{equation} \label{basiceq}
\langle 0 | A_i^\mu(x) | \pi^a(p) \rangle = {i p_\mu F_\pi
\delta_i^a  e^{i p \cdot x}\over (2\pi)^{3/2}\sqrt{2\, p^0}}
\end{equation}
where the $A_i$ are the currents associated with the
broken generators and the $\pi^a$ are the three pions,
and $F_\pi$ ($\simeq 184$ MeV in four dimensions)
is the phenomenological pion decay constant.
We will later see that the ``pion'' is merely required to be a
single-particle occupation number state, which transforms
as a (pseudo-)scalar under the Lorentz transformation;
the ``pion-decay constant'' will essentially stand for
a measure of the probability amplitude for a current
to interpolate between the vacuum and a one-particle occupation
number state.   The statements above may easily be generalized
to the breakdown of a symmetry $G\to H$, where $G$ and $H$ are global
symmetries associated with {\it underlying theory}, the ``pions'' resulting
from its (possibly unknown) dynamics of the underlying
theory.  There may be no a priori guide
to knowing what $H$ could be and how the pion-decay
constant(s) could be related to
the parameters of the underlying theory.  We will also
treat equations of the type given in eq.~(\ref{basiceq}) as the 
equation to define the pion decay constant matrix, in the case of
a generalized breakdown $G\to H$.  It will also not be necessary to 
refer to any generator as being broken or not;  in general, the number of
currents will be the dimension of the group of symmetries
and the number of ``pion'' states can also be as large
as the dimension of the group.  In this work, we shall show that all
pion-decay constants in the sense described above vanish
in $d=2$.  

At leading order, the identify eq.~(\ref{basiceq}) is reproduced, due to
the Lie-algebra isomorphism between $SU(2)\times SU(2)$ and $O(4)$, 
by the non-linear $O(4)$ sigma model lagrangian, ${\cal L}^{(2)}$, 
\begin{equation}\label{sigmamodelcurrent}
{\cal L}^{(2)}=-{1\over 2}{\partial_\mu \vec{\pi} \partial^\mu \vec{\pi}
\over \left(1+  \vec{\pi}^2/ F_\pi^2\right)}
\end{equation}
This yields the first term of the derivative expansion,
at $O(p^2)$, which is a term of $O(1/F_\pi^2)$.  
This follows from the expression for the Noether current
\begin{eqnarray}
& \displaystyle \vec{A}^\mu=F_\pi\left[\partial^\mu\vec{\pi}
{(1-\vec{\pi}^2/F_\pi^2)
\over (1+\vec{\pi}^2/F_\pi^2)}+{2\vec{\pi}(\vec{\pi}\cdot \partial^\mu
\vec{\pi})\over
F_\pi^2(1+\vec{\pi}^2/F^2)}\right]+... & 
\end{eqnarray}
associated with the invariances of the model, described in great
detail in Chapter 19 of Ref.\cite{sw2} (see also Ref.~\cite{gl2}).
It must be noted that the coupling constant of the sigma model 
defined by the equation abovis $1/ F_\pi^2$.  
Some additional remarks are in order:  
(a) the structure of the symmetry 
alone, and the assumption that Goldstone bosons are present in the spectrum 
dictate the form of ${\cal L}^{(2)}$ and is therefore valid in arbitrary
dimensions.   We note, however, that in the case of axial-vector symmetries
there is something of a conflict, since $\gamma_5$ is defined only in
even dimensions.  Our point of view is one where the underlying theory
is considered to be well defined in a given space-time dimension
and the corresponding effective lagrangian is constructed in that dimension
and then continued to $d\to 2$, and   
(b) the expression for the Noether current is obtained by
treating the Lagrangian around its perturbative vacuum, which
assumes that the interactions of the model do not alter the
spectral assumptions, viz., that the pions are present in the
spectrum as massless particles, and furthermore, and this is important
for our considerations, that the expansion parameter of the model
$1/F_\pi^2$ remains perturbative.

The power counting analysis corresponding to the low
energy expansion of the path integral is based on the observations that
(a) Goldstone bosons have only derivative couplings and each
derivative brings in a power $p$ of the momentum, (b)
that each loop in $d$ dimensions brings in the momentum $p^d$ powers
of momentum from $d^d p$ integral in the loop, and (c) that
each internal line brings in $p^{-2}$ powers of the momentum from
the $p^2$ term in the denominator of the propagator.  Thus
the presence of a vertices $V_i$ of order $d_i$, in an arbitrary diagram 
$\gamma$ with $L$ loops and $I$ internal lines, are associated with
the contribution to the generating functional of order $O(p^{O_\gamma})$, 
where 
\begin{equation}
O_\gamma=\sum_i V_i d_i + d L -2 I.
\end{equation}
One may eliminate $I$ through the use of the topological identity
\begin{equation}
L=I-\sum_i V_i+1,
\end{equation} 
to obtain
\begin{equation}
O_\gamma=\sum_i V_i (d_i-2) + (d-2) L +2 .
\end{equation}
This shows clearly why in $d=4$ at tree-level where there is
only one type of vertex of order 2 and no loops, the chiral
power counting produces contributions of $O(p^2)$, and why
the loops are suppressed by two powers of the momentum.
The argument above leads in general to a power-counting
that facilitates a low-energy expansion of Green's functions
in the powers of the momenta based on non-renormalizable lagrangians.
An attempt to work out the theory at one-loop
level leads to divergences, which are then absorbed in the
parameters of a new lagrangian ${\cal L}^{(4)}$.  The Green's
functions of the currents computed with this lagrangian are
then accurate at $O(p^4)$; it is also easy to see that
the expansion is also one is powers of
$F_\pi^{-2}$.  Note that the loops of ${\cal L}^{(2)}$
account for terms of $O(p^4)$. Setting $d=2$ shows that the loops 
are not suppressed~\cite{hl1} and thus the effective lagrangian 
analysis which start with only tree-level contributions cannot work 
in $d=2$.    In what follows, we shall show the apparent failure of
the loop-expansion can infact be traced to the absence of an expansion
parameter; the pion-decay constant itself will vanish and an expansion
in powers of $F_\pi^{-2}$ is ill-defined.    This result is derived from
an analysis of the underlying Wightman functions of field theory and
is independent of any lagrangian analysis.   Whereas it has been
recognized that the field parameterization independence of 
effective lagrangian techniqes~\cite{cwz} have
analogs from results in axiomatic field theory due to Haag~\cite{haag}, 
our results unfortunately do not provide such an analog in arbitary 
dimensions to the discussion of Ref.~\cite{hl1}, where the justification
for program of effective field theory is
carried beyond those of current algebra, and is provided
in terms of a direct analysis of the Ward identities of QCD via 
the generating functional technique.  

\setcounter{equation}{0}
\section{The Goldstone Mechanism Revisited}
The failure of the Goldstone mechanism in
2 dimensions was demonstrated by considering one of the proofs
and exhibiting the precise manner in which is fails,
which we briefly recapitulate.  In the notation of
Ref.~\cite{sc1},  consider a scalar field
theory described by a local scalar field $\phi$ and if
$j_\mu$ is a conserved current, then the scalar field defined by
\begin{equation}
 \delta \phi(y)=i\int d^{d-1}{\bf x} [j_0(x_0,{\bf x}),\phi(y)]  ,
\end{equation}
where the integration is over the spatial dimensions, is such that
its vacuum expectation value, for the Goldstone mechanism to succeed,
should differ from zero, i.e.,
\begin{equation}\label{colemancondition}
\langle 0 | \delta \phi (0) | 0 \rangle \neq 0.
\end{equation}
In 2 dimensions, however, it can be shown that this integral must vanish.  
This is related to the fact that the Wightman function:
\begin{equation}
\langle 0 | \eta(x) \eta(0) | 0 \rangle=\int {d^2k\over 2 \pi} e^{ik\cdot x}
\delta(k^2) \theta(k_0)
\end{equation}
fails to be well-defined, where $\eta(x)$ describes a free
massless scalar field satisfying the equation $\Box \eta=0$.   
In other words, it is not possible to define a free massless
scalar field theory in two dimensions, and that the Feynman propagator 
in two dimensions $\Delta_F(x-y)$ is also not well-defined.

Of the two proofs of the Goldstone mechanism in general that were proposed
in Ref.~\cite{gsw1},  the second proof of Ref.~\cite{gsw1}  
considered the effective action $\Gamma[\phi]$
and imposed on it the condition of invariance.   
Here we consider, in the notation of Ref.~\cite{sw2}, the case of the 
invariance of a field theory described by a set of hermitian fields (possibly
composite) $\phi_n$, subjected to the linear infinitesimal transformations
\begin{equation}
\phi_n(x) \to \phi_n(x) + g^\alpha (t_\alpha)_n^m \phi_m (x)
\end{equation}
where the $t_\alpha$ are the generators of the underlying symmetry
and the $g^\alpha$ are the parameters that parametrize an infinitesimal
transformation.  The symmetry is completely described the
by commutation relations of the algebra given by
\begin{equation}
[t_\alpha,t_\beta]=i f_{\alpha\beta}^\gamma t_\gamma
\end{equation}
where the $f_{\alpha\beta}^\gamma$ are the structure constants of the
algebra.  The set of conserved currents associated with
the symmetries $J^\mu_\alpha(x)$ are such that:
\begin{equation}
\partial_\mu J^\mu_\alpha(x)=0
\end{equation}
and the corresponding charges are
\begin{equation}
Q_\alpha=\int d^{d-1}{\bf x} J^0_\alpha(x)
\end{equation}
which satisfy the algebra
\begin{equation}
[Q_\alpha,Q_\beta]=i f^{\gamma}_{\alpha\beta} Q_\gamma.
\end{equation}
We also have the the transformation law:
\begin{equation}
[Q_\alpha,\phi^n(x)]=(t_\alpha)^n_m \phi^m(x),
\end{equation}
leads us to the consideration of the ``order parameter matrix''
of vacuum expectation values
\begin{equation}\label{opm}
C^m_\alpha=(-i) \langle 0 | [Q_\alpha, \phi^m(x)] | 0 \rangle
\end{equation}
The assumption of translational invariance
then implies the following statement for the effective potential
with $\Gamma[\phi]=-{\cal V}V(\phi)$, where ${\cal V}$ is the
space-time volume:
\begin{equation}
{\partial V(\phi) \over \partial \phi_n} g^\alpha(t_\alpha)_n^m \phi^m = 0.
\end{equation}
Taking a second derivative of this expression with respect to
the field, and the assumption that the potential is minimized
for $\overline{\phi}_n\neq 0$,
({\it viz,}
that the symmetry is spontaneously broken) then implies:
\begin{equation}
\sum_{n,m} {\partial^2 V(\phi) \over \partial \phi_n
\partial \phi_l} g^\alpha(t_\alpha)_n^m \overline{\phi}_m=0.
\end{equation}
In order for this equation to be satisfied, the matrix
\begin{eqnarray}
& \displaystyle {\partial^2 V(\phi) \over \partial \phi_n
\partial \phi_l} & \nonumber 
\end{eqnarray}
must have zero eigenvalues, whose eigenvectors
would be inverse propagators for massless scalar fields.
However, as noted earlier in $d=2$ the corresponding
propagators are ill-defined.  It then follows that
the assumption $\overline{\phi}_n=0$ we made in the first place is wrong 
and this implies that the Goldstone mechanism fails. 
This proof, however does not yield the desired result.   

A third proof of the Goldstone mechanism,
is the consequence of a Ward identity~\cite{gl2}.  This is shown
by considering the following time ordered product:
\begin{equation}
F_\alpha^{\mu m }(x)=\langle 0 | T J_\alpha^\mu(x) \phi^m(0) | 0 \rangle
\end{equation}
which satisfies the following equation for its divergence:
\begin{equation}
\partial_\mu F_\alpha^{\mu m}(x)=\delta(x_0) 
\langle 0 | [J_\alpha^0(x), \phi^m(0)] | 0 \rangle
\end{equation}
The assumption of current algebra that the
commutation rules hold at the local level, {\it viz.},
\begin{equation}
[J_\alpha^0(x),\phi^m(0)]=\delta^{(d-1)}({\bf x})[Q_\alpha,\phi^m(0)]=
-\delta^{(d-1)}({\bf x}) (t_\alpha)^m_n \phi^n(0),
\end{equation}
leads to the Ward identity
\begin{equation}
\label{inter}
\partial_\mu \langle 0 | T J_\alpha^\mu (x) \phi^m(0) | 0 \rangle =
-\delta^d(x) (t_\alpha)^m_n \langle 0 | \phi^n (0) | 0 \rangle.
\end{equation}
The Fourier transform of $F_\alpha^{\mu m}(x)$,
$\tilde{F}_\alpha^{\mu m}(p)$ then obeys the equation:
\begin{equation}
(-i) p_\mu \tilde{F}_\alpha^{\mu m}(p)=-(t_\alpha)^m_n \langle
0 | \phi^n (0) | 0 \rangle = i C^m_\alpha
\end{equation}
From Lorentz invariance, $\tilde{F}_\alpha^{\mu m}(p)$ must
have a pole at $p^2=0$.  This implies the existence
of massless bosons, the pions $\pi^a$.  The number of pions
depends on the characteristics of the order parameter matrix,
{\it viz.} the residual symmetry.  One may also define
the following matrix elements 
\begin{small}
\begin{eqnarray}
& \displaystyle \langle 0 | J^\mu_\alpha(x) | \pi^a(p) \rangle =
{i p^\mu F^a_\alpha e^{ip\cdot x}\over (2\pi)^{(d-1)/2}\sqrt{2\, p^0}}, \, 
\langle \pi^a(p) | \phi^m (x) | 0 \rangle = {G^{ma}
e^{-ip\cdot x}\over (2\pi)^{(d-1)/2}\sqrt{2\, p^0}}, &
\end{eqnarray}
\end{small}
[we have reserved the symbol $F_\pi$ for the case of QCD]
which is then related related to the order parameter matrix eq.(\ref{opm}) 
\begin{equation}
\sum_{a=1}^{N_\pi} F^a_\alpha G^{ma}=C^m_\alpha
\end{equation}
A crucial assumption that goes into the construction above,
is the existence of one-particle occupation number states,
which furnish a representation for the Lorentz group, and
must transform as (pseudo-)scalars, in order to not break Lorentz
invariance.  [The absence of such states would not allow us
to even define the matrix elements of interest to us, and
$F_\pi$ in particular would remain undefined.]   This assumption
can then be translated to a field renormalization prescription 
by which $G^{am}\neq 0$ can be held fixed.
This proof, however, breaks down in $d=2$ since in the intermediate
step, eq. (\ref{inter}), the vacuum expectation of the commutator
encountered with arbitrary scalar operator $\phi^m$, where
$\phi^m$ is only required to transform as a Lorentz scalar, and 
vanishes due to a generalization of eq.(\ref{colemancondition}).  
This must imply that the constants $C^m_\alpha$ for the choice must be zero.

As an illustration, let us consider the case in $d>2$ of the $O(N)$ linear model broken
down to $O(N-1)$ through the vacuum expectation value of the direction
$(1,0,...0)$ of a scalar multiplet transforming in 
the vector representation of $O(N)$.  It has been shown in Ref.~\cite{sw2}
that for this case, the following maybe established:
$F_{b}^a=\delta_{b}^a F$,  $G^{ab}=\delta^{ab} G, \, a,b=2,...,N$, and that
all other pion-decay constants and field renormalization constants vanish.   
[here there is only one pion-decay constant].  The only
non-vanishing $C^m_\alpha$ then are those of the structure
$FG\delta^b_a$.  However, for $d=2$, these too would have to vanish.
Note once more that a field renormalization prescription has to be available
in order to render the $|\pi^a(p)\rangle$ to be well-defined.  Thus,
since $G$ is held fixed, we must therefore have, $F=0$.
In the above, the discussion may as well have been for QCD,
with the scalar field that acquires a vacuum expectation value
replaced by an appropriate composite operator therein, and
upon further specializing the discussion to the $O(4)$ case 
(alternatively the $SU(2)\times SU(2)$ case, with the currents 
$J_\alpha$ now identified with the axial vector currents $A_i$) 
we see that
\begin{equation}
F_\pi(d=2)=0
\end{equation}
as required.    This proof, however, cannot be extended for
the general case, where little is known about the structure
of the entries of the pion-decay constant matrix and the 
field renormalization prescription matrix elements $G^{am}$.

A final proof we consider that avoids taking a recourse to
the field renormalization prescription arises when we consider
the spectral representation for products of currents:
\begin{eqnarray}
& \displaystyle
\sum_N \delta^d(p-p_N)\langle {\rm 0} | J^\mu_\alpha (0) | N \rangle
\langle {\rm 0} | J^\nu_\beta (0) | N \rangle^* =(2\pi)^{1-d}\theta(p_0)
 & \nonumber \\
& \displaystyle \times \left[\left(\eta^{\mu\nu}-p^{\mu}p^\nu/p^2\right)
\rho^{(1)}_{\alpha\beta}(-p^2)+p^\mu p^\nu \rho_{\alpha\beta}^{(0)}(-p^2)
\right], & 
\end{eqnarray}
where the $\rho^{(i)}_{\alpha\beta},\, i=0,1$ are spectral densities.
Taking the Fourier transform and using the completeness of
the states $|N\rangle$, this can be written as:
\begin{eqnarray}
& \displaystyle \langle 0| J^\mu_\alpha (x) J^\nu_\beta (0)|0 \rangle=
\int d\mu^2 & \nonumber \\
& \displaystyle \times \left[ \eta^{\mu\nu} \rho^{(1)}_{\alpha\beta}(\mu^2)
-\left( \rho^{(0)}_{\alpha\beta}(\mu^2) + \rho^{(1)}_{\alpha\beta}(\mu^2)/
\mu^2\right) \partial^\mu \partial^\nu\right] \Delta_+(x;\mu^2), &
\end{eqnarray}
where
\begin{equation}
\Delta_+(x;\mu^2)={1\over (2\pi)^{d-1}} \int d^{d}p \theta(p_0)
\delta(p^2+\mu^2) \exp(i p\cdot x)
\end{equation}
It may then be easily shown, using current conservation (viz., only
for non-anomalous currents) that
\begin{equation}\label{rhoequation}
\rho^{(0)}_{\alpha\beta}(-p^2)
=\delta(-p^2) \sum_a F_{\alpha}^a  F^{*a}_{ \beta}.
\end{equation}
Indeed, it has been pointed out in Ref.\cite{sc1} that a delta-function
singularity is forbidden on the light-cone in momentum space in $d=2$ for 
scalar fields,
which in turn implies that the matrix $K$ whose
elements are given by $\sum_a F_{\alpha}^a F^{*a}_{\beta}$
vanishes identically\footnote{
Constrast this with, in the notation of Ref.~\cite{sc1},
$\int d^2 x \exp^{ik\cdot x} \langle 0 | j_\mu(x)\,
\phi(0)|0\rangle=\sigma k_\mu \delta(k^2) \theta(k_0)+
\epsilon_{\mu\nu}k^\nu \rho(k^2) \theta(k_0)$, where $\sigma$ is
a numbers and $\rho$ is a scalar distribution. Here
the first term is well-defined due to the taming of the infra-red
singularity residing in the delta-function by 
$k_\mu$.  As a result, the proof that $\sigma$ must vanish,
is non-trivial~\cite{sc1}.}, in order to render
eq.~(\ref{rhoequation}) meaningful.  
The matrix $K$ is real symmetric due to the
vanishing of the commutator $
\langle 0|[J^\mu_\alpha (x), J^\nu_\beta (0)] |0\rangle$
for space-like separations, and furthermore, the one-particle basis
$|\pi^a(p)\rangle$ can always be rotated in such a manner that
the pion-decay matrix whose elements
$F_\alpha^a$ may be brought to a real diagonal form~\cite{hl1}  such that 
$F^{a}_{\alpha}=\delta_{\alpha}^a F_{(\alpha)}$. 
In this basis, it is easy to see that 
\begin{equation}
\rho^{(0)}_{\alpha\beta}(-p^2)
=\delta(-p^2)  F_{(\alpha)}  F_{ (\beta)} \delta^{\alpha\beta}.
\end{equation} 
At this point, we shall pick this basis and
it is then, a straightforward exercise to demonstrate
that the $F_{(\alpha)}$ all vanish.  This proves our proposition.
 
What can then be said about the non-linear sigma model and the
currents of the non-linear sigma model in the light of the above?  
In the case of the $O(N)$ non-linear sigma model, for $d=2+\epsilon$,
it is known that the model exhibits a phase transition~\cite{bls,bz}.  
A weakly coupled phase which
is one where the system is in the spontaneously broken phase; it is
precisely this phase which is required to reproduce, say 
eq.~(\ref{sigmamodelcurrent}) for $N=4$ and serves
as the effective theory for an underlying theory with Goldstone bosons.
Since the coupling constant of the non-linear sigma model is $F_\pi^{-2}$,
the weak coupling phase is one the in which $F_\pi^2$ cannot approach zero.
A second phase exists, which is a strongly coupled one with a bound
state $\sigma$ particle which is mass degenerate with the ``pions,'' 
and the model can no longer be the effective low energy lagrangian of 
QCD in $d=2+\epsilon$.   The collective modes of the resulting
model have been extensively studied, and the model now realizes
the complete $O(N)$ symmetry linearly~\cite{bls}.  In $d=2$, only the strongly
coupled phase is realized in accordance with our findings.
Another manner in which these results can be qualitatively understood
is that loops are unsuppressed, and even if a classical minimum is
chosen in the linear sigma model which breaks the symmetry, the effective
potential restores the symmetric minimum~\cite{cjp}.

\setcounter{equation}{0}
\section{Underlying theories in 2 dimensions}
It has been known for sometime that gauge theories in $d=2$
exhibit interesting and important properties.  The best studied
example is the Schwinger model~\cite{aar}; massless QED is $d=2$ with
one flavor which exhibits confinement, charge screening and chiral
symmetry breaking, associated with the $U(1)\times \tilde{U(1)}$
symmetries.  The symmetries of interest to us in the present
work, are those associated with $N_f$ flavors, and in particular
the $SU(N_f)_L\times SU(N_f)_R$ symmetry, which remains
unbroken as expected~\cite{bsrs}.  For the case of two flavors, the
theory has been again been recently studied~\cite{dsv}
 in somewhat greater detail and
the interactions associated with the so-called ``analog pions''
and ``analog sigma'' that realize the $O(4)$ symmetry that
is associated with this flavor symmetry.   The results
here remain in accordance with our findings.

Another interesting class of models are those based on $SU(N_c)$
gauge symmetry, with the fermions transforming in the fundamental
representation and with $N_f$ flavors.  
These provide a category of underlying theories
that reflect our own, viz., QCD.  These models have been widely
studied in the context of non-abelian bosonization~\cite{frishman}.  The limit
$N_c\to \infty, \, N_f=1$ is the well-known 't Hooft model~\cite{thooft}
which has no spin-zero bosons.  
The latter theory has been solved exactly in the
large $N_c$ limit and does not warrant a low energy analysis.
A massless, multi-flavor 
version of this model has been considered with $N_c$ fixed and
$N_f\to \infty$~\cite{armoni}, whose spectrum simplifies to one that is
$N_c^2-1$ copies of the abelian spectrum.  Here, it is clear
that the $SU(N_f)_L\times SU(N_f)_R$ symmetry remains explicitly
unbroken.


\setcounter{equation}{0}
\section{Conclusions}
A detailed analysis of the failure of the Goldstone mechanism
in $d=2$, first encoded by the Coleman and Mermin and Wagner theorem,
reveals through the analysis of the Ward identities and spectral
representations of the correlation functions of
currents, that the phenomenological pion-decay constant(s) 
must vanish.  In this manner, the would be effective lagrangian whose
structure is fixed by the symmetries and the reciprocal of the pion-decay constant,
for $d=2+\epsilon$ goes into a strongly coupled 
(and indeed for $\epsilon=0$ exists only in a symmetry restored)
phase, thereby vindicating the heuristic power
counting argument that says that effective lagrangians must
fail as we near $d=2$ for theories with pions.  Our work shows
the latter can essentially be justified on formal grounds in
field theory.  Our work is also consistent with results established
in the past concerning global flavor symmetries in exactly
solvable models, including the multi-flavor Schwinger model
as well as large $N_f$ $QCD_2$.

\bigskip

\noindent{\bf Acknowledgments:}
The hospitality of the International Centre for Theoretical Physics,
Trieste, Italy and of the Institut f\"ur Theoretische Physik,
Universit\"at Bern, Bern, Switzerland where part of this work
was completed is gratefully acknowledged.  
It is a pleasure to thank  N. D. Hari Dass, P. Hasenfratz, H. Leutwyler, 
P. Minkowski, S. Naik, V. P. Nair, A. D. Patel, S. Randjbar-Daemi,
D. Sen, C. S. Shukre and K. Sridhar for discussions.
I thank D. Sen for useful remarks on the manuscript.

\end{document}